\documentclass[final,3p,times,twocolumn]{elsarticle}

\usepackage{graphicx}

\usepackage{amssymb}
\usepackage{amsthm}

\journal{Physica A}

\begin{document}

\begin{frontmatter}



\title{Stability in a population model without random deaths by the Verhulst factor}


\author[nip,uplb]{C.M.N. Pi\~nol}
\ead{cpinol@nip.upd.edu.ph}

\author[nip]{R.S. Banzon\corref{cor1} \fnref{now}}
\ead{rbanzon@nip.upd.edu.ph}
\cortext[cor1]{Corresponding author}
\fntext[now]{Present address: National Institute of Physics, University of the Philippines Diliman, Quezon City 1101, Philippines; Tel: +632-920-9749, Fax: +632-9280296}

\address[nip]{National Institute of Physics, University of the Philippines \\ Diliman, Quezon City 1101, Philippines}
\address[uplb]{Institute of Mathematical Sciences and Physics, University of the Philippines \\ Los Ba\~{n}os, Laguna 4031, Philippines}

\begin{abstract}
A large amount of population models use the concept of a carrying capacity. Simulated populations are bounded by invoking finite resources through a survival probability, commonly referred to as the Verhulst factor. The fact, however, that resources are not easily accounted for in actual biological systems makes the carrying capacity parameter ill-defined. Henceforth, we deem it essential to consider cases for which the parameter is unnecessary. This work demonstrates the possibility of Verhulst-free steady states using the Penna aging model, with one semelparous birth per adult. Stable populations are obtained by setting a mutation threshold that is higher than the reproduction age.
\end{abstract}


\end{frontmatter}


\section{Introduction}
\label{Introduction}
The evolution of a population can be described using the difference equation

\begin{equation}\label{logistic}
N_{t+1} = rN_t\bigg(1 - \frac{N_t}{K}\bigg)
\end{equation}

\noindent where $N_t$ is the population at time $t$, $r$ is the intrinsic relative growth rate, and $K$ is the carrying capacity. This logistic type of growth was first introduced by Verhulst in the mid 1800s \cite{murray}. In an environment where resources are finite, populations are bounded. The equivalent differential equation leads to a final state which is a fixed point. Discrete counterparts, such as the one presented above, result in bifurcations, limit cycles and chaos \cite{murray,bernardes}. An in-depth discussion of the more complex set of solutions obtained by considering various aspects of the Verhulst equation may be found in \cite{ausloos-dirickx}.

The concept of a carrying capacity is commonly incorporated in population dynamics models. Limitations on food supply, space and other necessities suppress growth. Organisms compete in order to stay alive. Each subject to a survival probability, $V_t=1-N_t/K$, also known as the Verhulst factor \cite{moss,malarz&co}. The main motivation for its introduction is to make certain that population sizes are kept finite \cite{martins,makoweic,smoliveira&co,pekalski}. This parameter may be applied as a constraint for simulated populations in order to maintain their size within computer limits. However, the carrying capacity is not constant \cite{heilig}. An environment's actual ability to support life highly depends on prevailing knowledge and technologies \cite{simon}. It is not easily determined \cite{cohen}, neither is it well defined.

We choose to consider cases for which the carrying capacity is not essential. Because resources are not explicitly accounted for in many observed systems, we believe this case can be of great relevance to biological population studies.

\section{The model}
\label{The model}
The Penna model \cite{penna} is a popular technique for simulating aging populations \cite{smoliveira}. Individual characteristics are represented by a string of binary numbers of length $L$. Over time, mutations accumulate thereby modifying an organism's viability, survival and fertility \cite{puhl&co}. For each year added to an individual's age, one bit in the genome is read. Active genetic traits, therefore, are located at bit numbers (locus) less than or equal to the current age. Zeroes correspond to healthy genes and ones are bad or mutated genes. An individual dies a genetic death when it reaches old age ($age=L$) or when the number of expressed harmful mutations equals the threshold value, $T$. Here, only bad mutations are considered because helpful ones, in comparison, are very rare \cite{penna}. Reproductive maturity is achieved at age $R$. Starting from this age until death, individuals generate $B$ offsprings per time interval, where $B$ is the birth rate. In asexually reproducing populations, newborns copy the genes of the parent (both active and inactive) and acquire $M$ additional mutations. The mutation rate, $M$, can take on any value but is usually set to one since expression of new mutations is not very frequent in nature \cite{malarz}.

To include the effects of finite resources, individuals are subject to the same survival probability regardless of the quality of genome, fitness and age - VA implementation. It has been argued \cite{dabkowski} that older individuals should be more adaptable and, thus, have greater chances of survival. Furthermore, it was added that due to the Gompertzian behavior of the original model, limiting the number of births is enough to avoid exponential growth. Hence, an alternative approach was proposed applying the Verhulst factor to newborns only (the VB implementation). For lack of a vital role in population dynamics \cite{martins}, efforts have been made to eliminate the random deaths imposed by the Verhulst factor. In \cite{makoweic}, a Verhulst-free Penna model was implemented on a square lattice, where each lattice site can accommodate only one individual at a time. The abstract limit on the capacity of the environment is, thus, given by the size of the grid. Another study \cite{mossdeloiveiraetal} modified the model such that for every death, an adult is randomly chosen to produce one offspring (a replacement). This procedure keeps the total population size constant.

In this work, we modify the 32-bit Penna model, making it Verhulst-free. Deaths are now due only to genetics - old age and mutation accumulation. To limit growth, we allow individuals to breed only once in their lifetime (at age $R$). This behavior is observed in semelparous populations such as the mayflies and the Pacific salmon \cite{pennaoliveira1,pennaoliveira2,ortmanns}. The simulation starts with $N_0=20000$ perfect individuals (no bad genes) and runs for 3000 iterations. The choice of evaluating deaths before births limits the maximum reproduction age value, $R_{max}=L-1$.

\section{Analysis of results}
\label{Analysis of results}

\begin{figure}[ht!]
\centering
{\includegraphics[width=3in,angle=0]{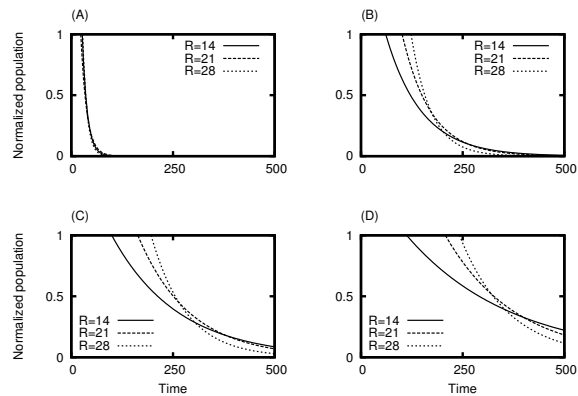}}
\caption{Exponential decay curves, $N/N_{max}=e^{-\alpha t}$, associated with $T \le R$: (A) $T=1, \alpha_{ave}=0.081$, (B) $T=5, \alpha_{ave}=0.015$, (C) $T=8, \alpha_{ave}=0.011$ and (D) $T=10, \alpha_{ave}=0.006$. Extrapolation shows that extinction begins at later times for populations with higher reproduction ages, $R$.}
\label{fig1}
\end{figure}

In the $B=1$ limit, the behavior of the resulting population is dependent on the relative value of the threshold parameter and the reproduction age. We observe exponential decrease when $T \le R$. Stability, on the other hand, is achieved by setting $T>R$. The populations presented hereafter are normalized about the maximum value obtained for each parameter set.

\subsection{Extinction cases}

Extinction has two causes - mutational meltdown and overpopulation \cite{malarz07}. The latter results from a sudden fluctuation in population size beyond the carrying capacity. But without the Verhulst factor (no stochasticity), the only possible causes of death are old age and mutation accumulation. At low $T$, selection favors deaths at younger ages in order to keep the gene pool clean. Thus when $T \le R$, a good fraction of the population dies prior to reproductive maturity. With births limited to one offspring per individual, the number of newborns is not enough to compensate for losses due to bad mutations. The population dies of mutational meltdown.

\begin{figure}[ht!]
\centering
{\includegraphics[width=3in,angle=0]{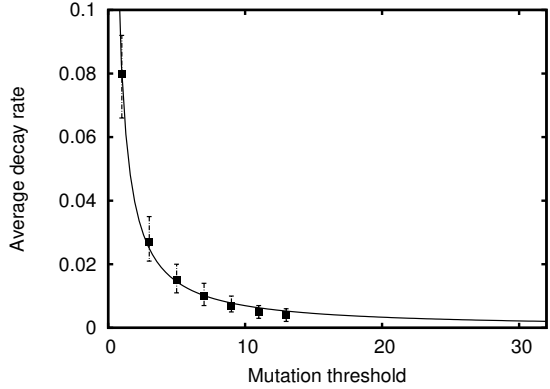}}
\caption{Average decay rate associated with $R=14,21$ and 28. The rate of extinction decreases with increasing tolerance for harmful mutations.  The relationship is a power law, $y=0.08x^{-1.06}$.}
\label{fig2}
\end{figure}

Best-fits show delays in the onset of extinction for those that reproduce late in life, large $R$ (Fig.~\ref{fig1}). The rate of decay is effectively slowed down by an increased tolerance for bad mutations. Regression analysis yields a power law relationship between the mutation threshold and the average decay rate (Fig.~\ref{fig2}).

\subsection{Nonzero steady states}

\begin{figure}[ht!]
\centering
{\includegraphics[width=1.5in,angle=0]{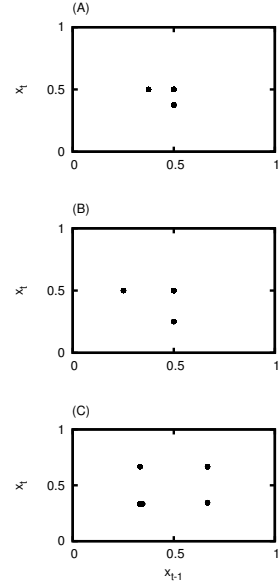}}
\caption{First return maps of stable populations resulting from $B=1,T=15$: (A) $R=4$, (B) $R=8$ and (C) $R=12$. These correspond to the last 300 iterations of the simulation, at which time the populations are already at steady state.}
\label{fig3}
\end{figure}

Stable populations were verified using first return maps \cite{chaos}. Consider the sequence $x_0,x_1,x_2,x_3 \ldots x_{t-1}$ and $x_t$. The first return map is a plot of consecutive entries of this sequence, i.e., $(x_0,x_1)$, $(x_1,x_2)$, $(x_2,x_3)$, \ldots $(x_{t-1},x_t)$. Fig.~\ref{fig3} presents the return maps of the steady state populations associated with $R=4,8$ and 12 when $B=1$ and $T=15$. The number of fixed points reflect periodic variations in the time series. For these steady states, we find that the magnitude of the fluctuations in total population size increases with $R$. 

\begin{figure}[ht!]
\centering
{\includegraphics[width=1.5in,angle=0]{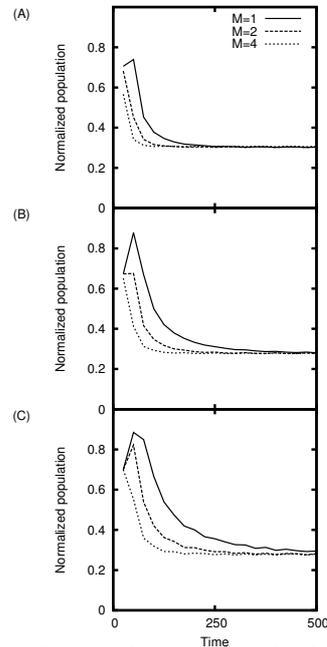}}
\caption{Normalized populations associated with $B=1$, $T=10$: (A) $R=3$, (B) $R=6$ and (C) $R=9$. Steady state is achieved faster for larger values of the reproduction age, $R$, and mutation rate, $M$.}
\label{fig4}
\end{figure}

To minimize fluctuations, running averages were obtained for the time series. This was done using intervals of 50. Variations in the transient part of the curves (Fig.\ref{fig4}) indicate differences in saturation time. The slope from the highest value to the onset of steady state becomes steeper as $R$ increases. Hence, those with higher reproduction ages take longer to equilibriate. Similar to the original Penna model results \cite{penna}, steady state is achieved faster by allowing more mutations, that is, by increasing $M$.

\begin{figure}[ht!]
\centering
{\includegraphics[width=3in,angle=0]{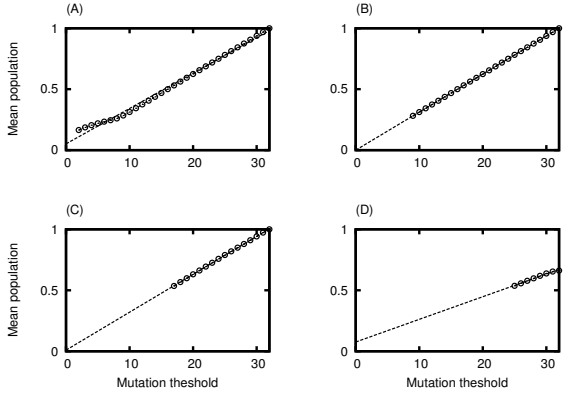}}
\caption{Mean normalized populations (circles) obtained from different threshold values and reproduction ages. Linear fits (dashed lines) yield (A) $y=0.029x+0.052$ for $R=1$, (B) $y=0.031x$ for $R=8$, (C) $y=0.031x+0.009$ for $R=12$ and (D) $y=0.019x+0.076$ for $R=24$.}
\label{fig5}
\end{figure}

Fig.~\ref{fig5} shows the variation of the mean population with threshold, for a given $R$. This corresponds to the average population size of the last 300 iterations of a particular simulation run. At these times, the simulated populations have already achieved steady state. Note that the mutation threshold can have values from 1 to $L$, the bit-string length. However, since stability is achieved only when $T>R$ (for the $B=1$ case), the number of $T$ values that result to nonzero steady states decreases as $R$ increases. We find that the total population scales with threshold. Considering all possible $R$ values (from 1 to $L-1$), we obtain linear fits that have slopes ranging from 0.015 to 0.035. 

The steady state associated with $T=L$ (or $M=0$) takes advantage of the finite bit-string characteristic of the Penna model. Following the derivation in \cite{pennaoliveira1}, we let $N(a,t)$ be the number of individuals with age $a$ at time $t$. The limit on the length of the string forces deaths at age $L$. In the absence of the other death factors, the discrete time evolution of the population is described by

\begin{equation}\label{individuals}
N(a+1,t+1) = \left\{ \begin{array}{ll}
N(a,t) & \textrm{$1 \le a < L-1$}\\
0 & \textrm{$a \ge L-1$}
\end{array} \right.
\end{equation}

\noindent and

\begin{equation}\label{babies}
N(0,t+1) = B N(R,t+1).
\end{equation}

\noindent At steady state, the average number of newborns is constant, i.e., $N(0,t)=N(0)$. From the equations above, we have

\begin{equation}\label{parents}
N(R,t+1) = N(0,t-R+1) = N(0).
\end{equation}

\begin{figure}[ht!]
\centering
{\includegraphics[width=3in,angle=0]{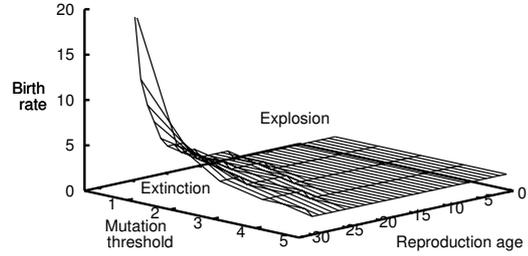}}
\caption{Population explosion curve associated with the Verhulst-free Penna model. Under the curve, behavior other than exponential decay is observed only if $B=1$ and $T > R$ (nonzero steady state). All threshold values not included in the figure result in exponentially increasing populations whenever $B>1$.}
\label{fig6}
\end{figure}

\noindent Using (\ref{parents}) in (\ref{babies}), it is straightforward to show that in the no mutation case ($T=L$ or $M=0$), stability is achieved only when $B=1$. No other nonzero steady states were found. Beyond $B=1$, populations either increase indefinitely or tend to zero (Fig.\ref{fig6}).

It has been shown \cite{martins-oliveira} that asexual populations have at most $T-1$ mutations before the minimum reproductive age. For semelparous species, all bits above age $R$ are set to one. In \cite{pinol-banzon}, we have demonstrated that this is not always the case. As $T$ increases, the strength of selection decreases and more mutations are preserved in the genome. When $T>R$, all bits at steady state contain defective genes. The process of mutation accumulation is very fast in asexual populations. To slow down the process, one may consider introducing recombination, as in sexual populations, or allowing positive mutations.

\section{Summary and conclusion}
\label{Summary and conclusion}
It is common practice in population models to attribute stability with the concept of a carrying capacity. Verhulst-free cases generally lead to Malthus catastrophe - exponential growth or decay. The carrying capacity, however, is an ill-defined parameter \cite{heilig,simon,cohen}. It is for this reason that we explore cases for which the concept is unnecessary, that is, to find stable populations without having to impose a Verhulst factor.

Within the framework of the Penna model, the artificial cap on the individual lifespan imposed by the bit-string length and the semelparous limit on reproduction make possible the Verhulst-free implementation.

Population explosion and extinction are still observed in many of the Verhulst-free cases. We find those associated with $B=1$ and $T>R$ result in nonzero steady states.








\end{document}